\documentclass[twocolumn,showpacs,preprintnumbers,amsmath,amssymb]{revtex4}
\usepackage{graphicx}
\usepackage{dcolumn}
\usepackage{bm}
\begin{document}

\title{Tails of the Density of States in a Random Magnetic Field}

\author{Riccardo Mazzarello$^{1.2}$, Stefan Kettemann$^{3}$, and Bernhard Kramer$^{3}$}

\affiliation{$^{1}$International School for Advanced Studies (SISSA), I-34014 Trieste, Italy\\ 
$^{2}$International Centre for Theoretical Physics (ICTP), I-34014 Trieste, Italy\\
$^{3}$I. Institut f\"ur Theoretische Physik, Universit\"at Hamburg, D-20355 Hamburg,  Germany}

\date{\today}

\begin{abstract}
  We study the tails of the density of states of fermions subject to a random
  magnetic field with non-zero mean with the Optimum Fluctuation Method (OFM).
  Closer to the centres of the Landau levels, the density of states is found
  to be Gaussian, whereas the energy dependence is non-analytic near the lower
  bound of the spectrum.
\end{abstract}
\pacs{71.23.An, 73.43.Cd, 73.43.Nq}

\maketitle

The problem of a charged quantum particle constrained to move in a two
dimensional (2D) static random magnetic field (RMF) has attracted considerable
theoretical and experimental interest in the past few years. The model plays
an important role within the composite fermion picture of the fractional
quantum Hall effect \cite{HLR}. Furthermore, it is supposed to describe states
with spin-charge separation in high-$T_{\rm c}$ superconductors \cite{highTc}.
It is also relevant to the understanding of the properties of a two
dimensional electron gas (2DEG) in lattice-mismatched InAs/InGaAs
heterostructures in magnetic fields \cite{Hansen}. In the latter systems the
electron gas is non-planar due to the lattice-mismatched epitaxial growth.
When a uniform magnetic field $B$ is applied, the electrons experience an
effective inhomogeneous field perpendicular to the non-planar
2DEG \cite{Hansen}. In addition, a static RMF in 2D inversion layers can be
experimentally realized in several ways. One possibility is to use a type-II
superconductor with a disordered Abrikosov flux lattice in an external
magnetic field as the substrate for the 2DEG \cite{Geim}. Alternatively, a
magnetically active substrate such as a demagnetized ferromagnet with randomly
oriented magnetic domains may be used \cite{Mancoff}. Recently, static RMFs in
2DEGs were created by applying strong magnetic fields parallel to GaAs
Hall-bars decorated with randomly patterned magnetic films \cite{Iye}.

The most fundamental quantity for
understanding the electonic properties of a random system is the density of
energy levels. The standard method to estimate the density of states (DOS) is
to calculate the imaginary part of the trace of the
single-particle Green function by diagrammatic techniques. However, this
approach fails in the tails of an energy band where multiple scattering up to
infinite order has to be considered in order to  take into  account 
 correctly  the   effect of localisation of electrons.
Also, numerical approaches are bound to fail in the asymptotic tails since here the
eigenstates are determined by rare statistical fluctuations of the randomness.
Moreover, in the case of RMF, the perturbative approach is also fundamentally
problematic since one has to deal with the non-diagonal part of the Green
function, which is not gauge invariant. In addition, the calculation of the
Green function is plagued by infrared divergencies
\cite{Alt,Khv1,Ealt,Khv2,She} that are due to the long-range nature of the
correlations of the vector potential, even if the spatial correlations in the
RMF are short-ranged.
It has been suggested that  these divergencies are due  to the non-gauge-invariance of 
the Green function and therefore unphysical \cite{Ealt},
although, recently, a physical interpretation has been proposed \cite{She}. In order to avoid
such  difficulties, E.~Altshuler et al. \cite{Ealt,Ealt2} calculated the DOS of a
charge in a RMF using the semiclassical approximation.
This is valid when  the energy
$E$ is much larger than the cyclotron energy $\hbar \omega_c$  corresponding to the mean magnetic
field $B$.  Also, it should exceed $\Gamma$, the disorder induced width of the Landau Levels (LL). A
field theoretical approach has been used to determine the DOS in a RMF with
zero mean value near the band edge \cite{Khv1}. The tail of the DOS in a
system of randomly distributed flux tubes of fixed strength was considered
\cite{Schofield}. There are also several numerical studies of the spectrum
with different mean and correlation lengths \cite{Othsuki}. Recently,
mathematically rigorous results have been obtained \cite{Ueki,Nakamura}. In
particular \cite{Ueki}, upper and lower bounds for the logarithm of the
integrated DOS near $E=0$ of some simple Gaussian RMFs with zero mean values
have been estimated.
For RMFs with non-zero mean value, the limit when $E$ is smaller than  
$\hbar \omega_c$ and, more generally, the tails of the lower Landau bands 
have not been considered analytically so far.  

It is our purpose  to provide nonperturbative  results for
the DOS in the tails of the lower Landau bands, as broadened by a static
RMF.
We show that  the
Optimum Fluctuation Method (OFM)
 \cite{Lifshitz1,Lifshitz2},
 being  non-perturbative and free from divergencies,
 can be extended  to treat  this RMF problem.
  We consider non-interacting fermions
 in  a RMF with non-zero mean, $B+b(\mathbf{r})$, with $b(\mathbf{r})$ Gaussian
distributed. The Hamiltonian  
\begin{equation}
  \label{eq:hamiltonian}
 H=\frac{1}{2m_{\rm e}}\left(\mathbf{p}-\frac{e}{c}\mathbf{A}\right)^2\,. 
\end{equation}
has a sharp lower bound of the energy spectrum at $E=0$. We
concentrate on the energy region near the first Landau band.
Since the OFM is especially designed to grasp 
rare fluctuations, it allows us to calculate the energy- and  $B$-dependence of the leading terms of
the DOS.
The correlation function of the RMF is assumed as
\begin{equation}\label{2corrB}
\langle b(\mathbf{r}) b(\mathbf{r}^{\prime
  })\rangle=\beta(|\mathbf{r}-\mathbf{r}^{\prime}|).
\end{equation}
with $\beta(|\mathbf{r}-\mathbf{r}^{\prime}|)\rightarrow 0$ as
$|\mathbf{r}-\mathbf{r}^{\prime}|\rightarrow \infty$. The random field is
ergodic, i.e. the correlations between different regions decay to zero with
increasing distance. We also assume that $\beta(r)$ is characterized by a
single scale $r_{c}$, the correlation length of the RMF. The
probability density for a specific realization $b(\mathbf{r})$ with correlator
(\ref{2corrB}) is $P[b(\mathbf{r})]=\mathcal{N}\exp\{-S[b(\mathbf{r})]\}$,
where $\mathcal{N}$ is the normalization constant and $S[b(\mathbf{r})]$, the
action of the RMF, is
\begin{equation}\label{2actioB}
S[b(\mathbf{r})]=\int  
b(\mathbf{r})\beta^{-1}(\mathbf{r}-\mathbf{r}^{\prime }) 
b(\mathbf{r}^{\prime })d\mathbf{r}d\mathbf{r}^{\prime }.
\end{equation}
The kernel $\beta^{-1}(\mathbf{r}-\mathbf{r}^{\prime })$ has the property
\begin{equation}
\int d\mathbf{r}^{\prime}\beta^{-1}
(\mathbf{r}-\mathbf{r}^{\prime})
\beta(\mathbf{r}^{\prime}-\mathbf{r}^{\prime\prime})
=\delta(\mathbf{r}-\mathbf{r}^{\prime\prime}).
\end{equation}
For a $\delta$-correlated RMF, the action reduces to
$S[b(\mathbf{r})]=1/\beta_{0}\int b^{2}(\mathbf{r})d\mathbf{r}$. The
configurationally averaged DOS is
\begin{equation}\label{DOSaver}
\rho(E)= \int D b(\mathbf{r}) P[b(\mathbf{r})] \rho(E;[b(\mathbf{r})]).
\end{equation}

In the low energy tail of the
first Landau band states are expected to be localized near
strong, exponentially rare  fluctuations of $b(\textbf{r})$.
 Therefore, the average in Eq.  (\ref{DOSaver}) over all
configurations, yielding  states at energy $E$ is dominated by the most
probable realization of $b(\textbf{r})$.
The functional integral in Eq. 
(\ref{DOSaver}) can be evaluated using the saddle point approximation.
Furthermore, only fluctuations in which the lowest level $E_{0}$ is equal
to $E$ have to be taken into account since a configuration in which $E$
corresponds to an excited level is  less probable.
Within logarithmic accuracy
\cite{Lifshitz1,Lifshitz2,Zittartz,Friedberg,Houghton,Larkin81},
\begin{equation}\label{Lifshitz}
-\ln \rho (E)\sim\mathop{\mathrm{min}}_{b(\mathbf{r})}
 S[b(\mathbf{r})]|_{E_{0}[b(\mathbf{r})]=E}.
\end{equation}

 Before we present a rigorous treatment,
let us  first give an intuitive estimate for the DOS 
closer to the band center,  $\Delta E\ll\hbar \omega_{\rm c}$,
 where $ E =  \hbar \omega_{\rm c} - \Delta E $.
 For short-range RMF, $r_{c}\ll l_{B}$, with
$l_{B}=(\hbar c/e B)^{1/2}$ the magnetic length related to $B$, 
 an optimal configuration near the band centre is likely to be a
circular magnetic well with depth $\Delta b\ll B$.
The corresponding action is
 $S\approx
\pi\Delta b^{2} R^{2}/\beta_{0}$, 
 where $R$ is the radius of the well. 
 The ground state energy $E_{0}$ of a
charged particle in a circular
magnetic well of radius $R$, where 
 the magnetic field is different from the 
 constant magnetic field B,
 is known \cite{solimany}.
The radii $R_{\rm
  opt}$ of the  optimal wells with  lowest
action $S_{\rm min}(\Delta E)$, are proportional to $l_{B}$ and
independent on $\Delta E$. Moreover, the depths $\Delta
b_{\mathrm{opt}}\ll B$ of these wells are proportional to $\Delta E$ and
do not depend on $B$,
\begin{equation}
R_{\mathrm{opt}}\sim 1.6\, l_{B}
\propto B^{-1/2},\qquad \Delta E\propto\Delta b_{\mathrm{opt}}.
\end{equation}
The action of the optimum fluctuation with $E_{0}=E$ is then
\begin{equation}\label{actnearC}
S(\Delta E)\sim 
\frac{R_{\mathrm{opt}}^{2}\Delta b_{\mathrm{opt}}^{2}}{\beta_{0}}
\sim \frac{l_{B}^{2} \Delta E^{2}}{\beta_{0}}.
\end{equation}
From (\ref{Lifshitz}) and (\ref{actnearC}) we obtain
\begin{equation}\label{DOSnumberone}
\rho(E)\sim \exp\left(-\Delta E^{2}/\Gamma_{\delta,0}^{2}\right),
\end{equation}
with $\Gamma_{\delta,0}=\alpha \hbar e \beta_{0}^{1/2}/(m_{\rm e}c\, l_B)$,
where $\alpha\sim 1.5\cdot 10^{-2}$ is a numerical factor. The variance of
the Gaussian is thus proportional to $B$. Our simple arguments are expected to
be valid when the energetic distance from the center of the lowest LL fulfils
$\Gamma_{\delta,0}\ll\Delta E\ll\hbar\omega_{\rm c}$.
A completely analogous argument holds for the right tail of the first LL, near
the band centre. In this case the optimal fluctuations are magnetic circular
humps with height $\Delta b\ll B$ and radius $R\propto l_{B}$ and the leading
exponential term of the DOS in the right tail shows the same dependence on the
energy shift and $B$ as (\ref{DOSnumberone}).

With  long-range RMFs, the analysis of the DOS near the band centre
is simpler: the localization radius of a typical state of the order of
$l_{B}$ is much shorter than the radius of an optimal potential well. The
correlation length $r_{\rm c}$, and the energy $E$ of such a state is, 
 in leading order, equal to  the first LL energy in the total 
field $B-\Delta b$.
The energy shift is thus proportional to $\Delta b$, and  the RMF 
acts exactly like 
  a random electrostatic potential. Since for long-range RMFs the
radius of the well is the largest length scale, the probability distribution
(\ref{2actioB}) can be approximated as
\begin{equation}\label{Plongb}
P[\Delta b]\sim \exp \left(-\Delta b^2/\beta(0)\right).
\end{equation}
Hence, 
\begin{equation}\label{rholongb}
-\ln \rho(E)\sim\frac{\Delta E^{2}}{\Gamma_0^2}, 
\end{equation}
with $\Gamma_0=\hbar e \beta(0)^{1/2}/m_{\rm e}c$, and the exponent of the DOS
does not depend on $B$, as long as the inequality $l_{B}\ll r_{\rm c}$ is
fulfilled. Equation (\ref{rholongb}) is valid if $\Gamma_0 \ll \Delta E\ll
\hbar\omega_{\rm c}$.
Similar considerations are expected to yield a Gaussian DOS also in the tails
of higher Landau bands, in the regions $\Gamma_{n}\ll
|E-(n+1/2)\hbar\omega_{\rm c}|\ll\hbar\omega_{\rm c}$, where $\Gamma_{n}$ is
the width of the n-th LL. In these regions, the DOS resembles the one of
independent charged particles in a Gaussian electrostatic potential
\cite{Larkin81,Affleck}.

Due to the sharp  band edge at $E=0$
the DOS is expected to approach zero more rapidly  
 for energies $E\ll\hbar\omega_{\rm c}$. For short-range RMF, states
with arbitrarily small energies can  be obtained 
when they are  localized in regions of area $\mathcal{A}$,
inside which $b \approx 0$ and outside which $b\approx B$.
 The action of these fluctuations is
\begin{equation}\label{actionedge}
S\sim \mathcal{A}B^{2}/ \beta_{0}, 
\end{equation}
and  the ground state energy scales like the one in a potential well, 
$E\sim \hbar^{2}/2 m_{\rm e} A$.
Thus,   $1/A \propto E $  and the DOS becomes  a non-analytic function of $E$,
\begin{equation}\label{rhoedge}
\rho(E)\sim \exp\left(-K_{0}B^{2}/\beta_{0}E\right),
\end{equation} 
with $K_{0} \sim \pi\hbar^2/2m_e$.
The above picture is analogous to the argument used by Lifshitz to estimate
the tail of the DOS of a particle subject to a Poissonian random potential
generated by short-range, repulsive impurities in zero magnetic field
\cite{Friedberg,Lifshitz2,Leschke}. The argument holds for a Poissonian
distribution of magnetic fluxes, too \cite{Schofield}. 

Intuitively, for large correlation length, $r_{\rm c}\geq l_{B}$, there is a
right neighbourhood of $E=0$ such that
each corresponding  optimum well has a radius much larger than the correlation
length, $R\gg r_{\rm c}$. Equivalently, $E\ll \hbar^2/(m_{\rm e} r_{\rm
  c}^2)$. The larger the correlation length, the closer to the band edge the
energies of the optimal states must be in order to fulfil this inequality.
 Defining 
\begin{equation}
  \label{eq:betanull}
\beta_{0}\equiv \int
\beta(\mathbf{\xi})d^2\mathbf{\xi}\sim \beta(0)r_{\rm  c}^2,
\end{equation}
 the action of $b(\mathbf{r})$ is still given by Eq. 
(\ref{actionedge}) and 
the DOS  by Eq.  (\ref{rhoedge}).
Hence, for $E\rightarrow 0$, the DOS  becomes independent of the 
 correlation length  $r_{\rm c}$.

In order to obtain exact expressions for the DOS,  
we now derive the variational equations which determine the shape
of the optimal fluctuations and wave functions \cite{Houghton,Zittartz}. 
According to (\ref{Lifshitz}), we must search for
the maximum of the probability distribution Eq. (\ref{2actioB}) under the
constraint $E_{0}=E$. For the weaker constraint
\begin{equation}
\det \{H[b(\mathbf{r})]-E\}=0,
\end{equation}
or, equivalently, $E_{n}=E$ for some energy level $E_{n}$, the optimum
fluctuation $\bar{b}(\mathbf{r})$ of the RMF must satisfy
\begin{equation}\label{variatio}
\left. \!\!\!\!\int \!\beta^{-1}(\mathbf{s}-
\mathbf{s}^{\prime})\bar{b}(\mathbf{s}^{\prime })d\mathbf{s}^{\prime }+\mu 
\frac{\delta \det \{H[b(\mathbf{r})]-E\}}
{\delta b(\mathbf{s})}\right|_{b=\bar{b}}\!=0,
\end{equation}
where $\mu$ is a Lagrange multiplier. Using
\begin{equation}
\det \{H[b(\mathbf{r})-E\}=\exp(\mathrm{tr}\ln\{H[b(\mathbf{r})]-E\}),
\end{equation}
and assuming that the ground state energy $E_{0}[\bar{b}(\mathbf{r})]$ 
is equal to $E$, we find
\begin{equation}\label{variatio2}
\left. \int \beta^{-1}(\mathbf{s}-\mathbf{s}^{\prime})
\bar{b}(\mathbf{s}^{\prime
  })d\mathbf{s}^{\prime }+\mu^{\prime}(E) 
\frac{\delta E_{0}[b(\mathbf{r})]}{\delta 
b(\mathbf{s})}\right|_{b=\bar{b}}=0
\end{equation}
with $\mu'(E)\equiv\mu\prod_{n=1}^{\infty}(E_{n}-E)$. In the Coulomb gauge, we
can write the Hamiltonian as a function of $b(\mathbf{r})$ and
calculate $\delta E_{0}/\delta b(\mathbf{s})$. The
variational equation (\ref{variatio2}) yields
\begin{equation}\label{Bopt}
\bar{b}(\mathbf{s})=-\mu'(E) \frac{e}{c}\int
d^{2}\mathbf{s}^{\prime}\beta(\mathbf{s}-\mathbf{s}^{\prime})
\int d^{2}\mathbf{r}\;
\mathbf{j}_{0}\cdot\mathbf{a}_{\Phi}(\mathbf{r}-\mathbf{s}^{\prime})
\end{equation}
where
\begin{equation} 
\mathbf{a}_{\Phi}(\mathbf{r}-\mathbf{s})=\frac{1}{2\pi}
\frac{\hat z\times(\mathbf{r}-\mathbf{s})}{|\mathbf{r}-\mathbf{s}|^{2}}
\end{equation} 
and $\mathbf{j}_{0}=\Psi_{0}^{*}\mathbf{\Pi}\Psi_{0}/2m_{\rm e}$ is the
ground state current, $\mathbf{\Pi}$ is the kinetic momentum of
the particle.  Eq. (\ref{Bopt}), together with 
\begin{equation}\label{Hopt}
\frac{1}{2m_{\rm e}}\Pi^2\Psi_{0}=E\Psi_{0}
\end{equation}
determines the optimal magnetic field  $b(\mathbf{r})$ and
 ground state wave function $\Psi_{0}(\mathbf{r})$.

We have solved equations (\ref{Bopt}) and (\ref{Hopt}) iteratively both in the
case $r_{\rm c}\ll l_{B}$ and in the case $r_{\rm c}\gg l_{B}$ for RMFs with
Gaussian correlators
\begin{equation}\label{Gauscorr}
\beta(\mathbf{r}-\mathbf{r}^{\prime })=
\frac{\beta^{\prime}}{2 \pi r_{\rm    c}^{2}}\exp
\left(-\frac{|\mathbf{r}-\mathbf{r}^{\prime }|^2}{2 r_{\rm c}^{2}}\right).
\end{equation} 
Since the RMF distribution function is rotationally invariant,
circular symmetry is assumed. 

\begin{figure}
\begin{center}
\includegraphics[width=8.5cm]{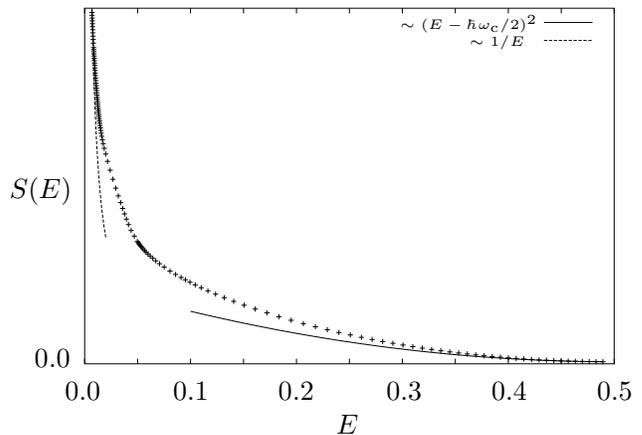}
  \caption{Behavior of the effective action
$S(E) = - \ln \rho (E)$  
 in a short-range RMF, $r_{\rm c}=0.1 l_{B}$.
 Energies are in units of $\hbar \omega_{\rm c}$ and $\Gamma_{0}\sim 10^{-4}\hbar\omega_{\rm c}$,
     $S$ is in arbitrary units.}
\label{fig:dosregion}
\end{center}
\end{figure}

For  short-range RMFs the result for the DOS  $\rho (E)$  is shown in
Fig.~\ref{fig:dosregion} as function of energy $E$, 
 where $S(E) = - \ln \rho (E)$.  
At the band edge we find indeed that the action of the optimum
fluctuation shows the characteristic behavior as a function of the energy 
$ E$, 
$ S \sim E^{-1}$
 while closer to the band center it changes to 
$ S\propto \Delta E^{2}$.
This reflects the physical origins of the corresponding typical wave
functions, and is consistent with the  qualitative arguments, given above.

\begin{figure}
\begin{center}
\includegraphics[width=8.5cm]{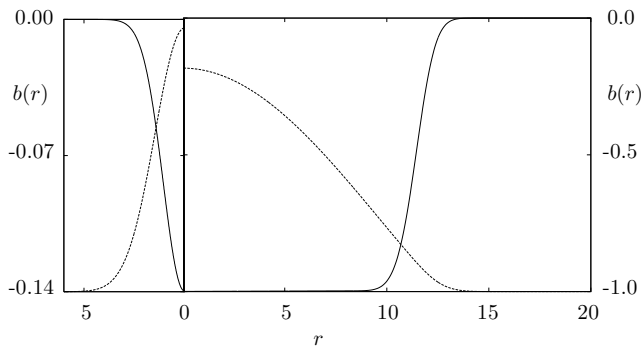}
\caption{The optimum magnetic wells $b(r)$ (solid line)
  and the corresponding ground state wave functions $\Psi(r)$ (dashed line) for correlation
  length $r_{\rm c}=0.1 l_{B}$. Left: closer to the centre of the band,
  energy shift \(\Delta E=3.6\cdot 10^{-2}\hbar \omega_{\rm c}\). 
  Right: near the band edge; ground state energy $E=1.8\cdot 10^{-2}\hbar 
  \omega_{\rm c}$. $r$ is in units of \(l_B\) and \(b(\textrm{r})\) in units of $B$.
\label{nearCLR}}
\end{center}
\end{figure}

Fig. ~\ref{nearCLR} shows   the optimal fluctuations near the
band centre and the band edge for  short-range fluctuations. Near the centre, 
the typical fluctuations are shallow wells, compared to $B$, with relatively 
steep walls. In the case of long-range fields,
 near the band centre, the radius of the ground state is much smaller 
than the size of the well.

In experiments some kind of random electrostatic potential is always
present. Let us
assume that the RMF and the random potential (RP) are independent random
quantities and that they are
both long-ranged. For a weak RP, $W(0)^{1/2}\ll \hbar e \beta(0)^{1/2}/(m c)$
(where $W(\mathbf{r})=\langle V(0)V(\mathbf{r})\rangle$), the RMF is dominant
in the tail at positive energies except for a narrow region close to $E=0$. 
At $E\ll \hbar\omega_{\rm c}$ the action of an optimal well
of the RP is $S_{RP} \sim (E-\hbar\omega_{\rm c}/2)^2/2W(0)\sim\hbar^{2}\omega_{\rm
  c}^2/4 W(0)$, whereas $S_{RMF}\sim \hbar^2\omega_{\rm  c}^2/(2mEr_{\rm c}^2\delta\omega_{\rm c}^2)$, 
where $\delta\omega_{\rm c}=e\beta(0)^{1/2}/mc$ and $r_{\rm c}$ is the correlation length of the RMF. 
The RMF will dominate on the RP if $S_{RMF}<S_{RP}$;
therefore, the exponent of the DOS will be proportional to $1/E$ if
$1\ll R^{2}/r_{\rm c}^{2}\ll\hbar^{2}\delta\omega_{\rm c}^2/W(0)$, with
$R^2\sim\hbar^{2}/mE$. 
Hence, the results presented in this paper 
break down at energy $E_{\rm  c}\sim W(0)/m r_{\rm c}^{2}\delta\omega_{\rm
  c}^2$. For large, negative energies, $E\rightarrow-\infty$, 
the RMF becomes irrelevant and the DOS is
purely classical, $-\ln \rho(E)=E^{2}/2W(0)$.

In conclusion, we have determined the density of states of a charged particle
in a spatially correlated, randomly varying magnetic field with non-vanishing
average. The latter provides Landau levels which  are broadened
into Landau bands by the RMF, well separated for sufficiently small disorder. In
the regions of the tails of the Landau bands which are accessible neither by
perturbative multiple scattering expansions nor by numerical calculations, we
have found that the average DOS is determined by the {\em typical}
configuration of the magnetic field. This is reflected in the energy
dependence of the effective action, $S(E)$, and the fact that the latter is
proportional to the logarithm of the DOS, $\rho(E)$, 
 which is found to be asymptotically
singular at the lower bound of the energy spectrum and becomes quadratic as a
function of the energy closer towards the centre of the band. In order to
determine the pre-exponential factor of the DOS, 
 one  integrates over the fluctuations of the
magnetic field around the saddle point configuration satisfying Eq.  (\ref{Bopt})
\cite{Houghton,Larkin81,Spies}.
As a final remark, we want to discuss briefly the relevance of our work to the CF description of the Fractional 
Quantum Hall Effect. Within this model, electrons are replaced by fermions experiencing a fictitious 
magnetic field proportional to the particle density in addition to the external one. In the presence of a random impurity potential, 
at the mean field level, the particle density is spatially inhomogeneous due to screening and the fictitious magnetic field has thus 
a spatially stochastic component \cite{HLR}. However, since the random impurity potential and the RMF are not independent random quantities, 
our theory cannot be straightforwardly applied to the CF model. This issue is subject to future work and will be published elsewhere.

We acknowledge useful discussions with M. Raikh, A. Struck, and E. Mariani.
This work has been supported by the EU via the RTN HPRN-CT2000-00144 and by 
the DFG and the DFG-Schwerpunkt ``Quanten-Hall-Effekt''.

\end{document}